\documentstyle[12pt]{article}
\begin{document}
\begin{titlepage}
\begin{centering}
{\large{\bf Classical moduli hair for Kerr\\
 black holes in String Gravity}}\\
\vspace{.7in}
{\bf P.Kanti and K.Tamvakis}\\
\vspace{.2in}%
Division of Theoretical Physics\\
Physics Department,University of Ioannina\\
Ioannina GR-451 10 ,GREECE\\
\vspace{.8in}
{\bf Abstract}\\
\vspace{.3in}

{\noindent
We compute the external moduli and dilaton hair to $O(\alpha')$  in the
framework of the one-loop corrected  superstring effective action for a
rotating black hole background.}\\
\end{centering}

\vspace{3.5in}
%

\par
\vspace{4mm}
\begin{flushleft}
IOA-316/95\\
hep-th/9502093
\end{flushleft}
\end{titlepage}

\par
Superstring $^{\cite{Green}}$ theory is our best existing candidate for a
consistent quantum theory of gravity which also has the prospect of unification
with all other interactions.Einstein's theory which has been very successful as
a classical theory of gravitation is incorporated in this more general
framework.However,Superstrings involve a characteristic length of the order of
the Planck scale and are expected to lead to drastic modifications of the
Einstein Action at short distances.These modifications arise either due to the
contribution of the infinite tower of massive string modes,appearing as
$\alpha'$ corrections,or due to quantum loop effects.An effective low energy
Lagrangian $^{\cite{Callan}}$ that incorporates the above, involving only the
massless string modes,can be derived from string theory using a perturbative
approach in both the string tension $\alpha'$ and the string coupling.The
relevant massless fields,apart from the graviton and other gauge fields,are the
dilaton,which plays the role of the field-dependent string coupling that
parametrizes the string loop expansion,
and the moduli fields that describe the size and the shape of the internal
compactification manifold.

In Einstein gravity,minimally coupled to other fields,the most general black
hole solution is described by the Kerr-Newman family of rotating charged black
hole solutions $^{\cite{Wald}}$.In agreement with the ``no-hair"theorem
$^{\cite{Israel}}$ at the classical level the only external fields present are
those required by gauge invariance.A qualitative new feature present in the
superstring effective action is the appearance of external field strength hair
for the axion and dilaton fields
$^{\cite{Gibbons}}$ $^{\cite{Perry}}$ $^{\cite{Campbell}}$
$^{\cite{Garfinkle}}$
  $^{\cite{Kaloper}}$ $^{\cite{Shapere}}$ $^{\cite{Olive}}$.The
tree level effective action has been calculated up to several orders in the
$\alpha'$-expansion.It turns out that there is no dependence on the moduli
fields
at tree level.The one-loop corrections to gravitational and gauge couplings
have
been calculated in the context of orbifold compactifications of the heterotic
superstring $^{\cite{Antoniadis}}$.It has been shown that there are no
moduli-dependent corrections to the Einstein term while there are non trivial
${\cal {R}}^2$-contributions appearing in a Gauss-Bonnet combination multiplied
by a moduli-dependent coefficient function.This term is subject to a
non-renormalization theorem which implies that all higher-loop moduli-dependent
${\cal {R}}^2$-contributions vanish.It is interesting to note the existence of
singularity-free $^{\cite{Tamvakis}}$ solutions of the field equations in a
Friedmann-Robertson-Walker background depending crucially on the presence of
the
Gauss-Bonnet term.

In the present short article we extend existing treatments
$^{\cite{Gibbons}}$ $^{-}$ $^{\cite{Olive}}$ of black hole solutions in string
gravity to iclude moduli fields.For simplicity we restrict ourselves to the
zero
charge case although the case of charged black holes is expected not to lead to
any extra complication. Our action is the low energy effective action derived
in
the context of orbifold compactifications of the heterotic superstring to
one-loop and $\alpha'$-order.We compute to $\alpha'$-order the moduli and
dilaton-hair together with the corresponding two axions hair.The result
although
expected from previous existing investigations without the moduli fields serves
to establish even better the qualitatively  new features of string gravity in
contrast to Einstein gravity characterized by the ``no-hair" theorem.In our
action we have not introduced any potential for the above fields although it is
likely that in the full quantum string theory such a potential and a (small)
mass is generated through non-perturbative effects.Nevertheless,if the black
hole size,or the distance from the black hole ,is small compared to their
inverse mass the solutions found are valid to a good approximation.

Let us consider the universal part of the effective action of any
four-dimensional heterotic superstring model  which describes the dynamics of
the graviton,the dilaton S and,for simplicity,the common modulus field T.At the
tree level and up to first order in $\alpha'$ it takes the form
\begin{equation}
S^{(o)} _{eff}  =  \int d^4 x \sqrt{-g} \left( {\frac {1} {2}} R + {\frac
{|DS|^2} {(S+ \overline{S})^2}} + 3 {\frac {|DT|^2} {(T+\overline{T})^2}}
  +  {\frac {\alpha'} {8}} (ReS) {\cal {R}}^2 _{GB} + {\frac {\alpha'} {8}}
(ImS)
{\cal {R}} {\tilde{\cal{R}}} \right)
\end{equation}
where
\begin{equation}
{\cal {R}}^2 _{GB} \equiv R_{\mu \nu \kappa \lambda} R^{\mu \nu \kappa \lambda}
- 4 R_{\mu \nu } R^{\mu \nu} + R^2
\end{equation}
and
\begin{equation}
{\cal{R}} {\tilde{\cal{R}}} \equiv {\cal{\eta}}^{\mu \nu \rho \sigma} R^{\kappa
\lambda} _{\;\;\;\mu \nu} R_{\rho \sigma \kappa \lambda}
\end{equation}
Note that\footnote{$\epsilon^{oijk} = -\epsilon_{ijk}$} ${\cal{\eta}}^{\mu \nu
\rho \sigma} \equiv {\epsilon}^{\mu \nu \rho \sigma} (-g)^{-1/2} $. We have
chosen units such that $k \equiv \sqrt{8 \pi G_N } \equiv 1$.

The one-loop corrections give a modulus dependence to the quadratic
gravitational terms that are of the form
\begin{equation}
S^{(1)} _{eff} = \int d^4 x \sqrt{-g} \left( \alpha'\Delta(T,\overline{T})
{\cal{R}}^2 _{GB} +\alpha' \Theta (T,\overline{T}){\cal{R}} {\tilde{\cal{R}}}
\right)
\end{equation}
The functions $\Delta(T,\overline{T})$ and $\Theta (T,\overline{T}) $ have been
derived in ref.$[12]$ and depend multiplicatively through a coefficient on the
supermultiplet content of the string model.Indroducing the notation
\begin{equation}
S \equiv ( e^{\phi} + i a)/{g^2} \quad ,\quad T \equiv e^{\sigma} + i b
\end{equation}
and reffering to $\phi$ as the dilaton, to $\sigma$ as the modulus,to
$a$ and b as the axions and to $g^2$ as the string coupling ,we can
write the effective one-loop, $O(\alpha')$ Langrangian as
\begin{eqnarray}
{\cal{L}}_{eff} & = & {\frac {1} {2}} R + {\frac {1} {4}} {({\partial} _{\mu}
\phi)}^2 + {\frac {1} {4}} e^{-2 \phi} {({\partial} _{\mu} a)}^2 +
{\frac {3} {4}} {({\partial} _{\mu}\sigma)}^2 + {\frac {3} {4}} e^{-2 \sigma}
{({\partial} _{\mu} b)}^2 \nonumber\\ & + & \alpha' \left( {\frac {e^{\phi}} {8
g^2}} + \Delta \right) {\cal{R}}^2 _{GB} +  \alpha' \left({\frac {a} {8 g^2}} +
\Theta \right){\cal{R}} {\tilde{\cal{R}}}
\end{eqnarray}

The equations of motion resulting from $(6)$ are four equations for the scalar
and pseudoscalar fields
\begin{equation}
{\frac {1} {\sqrt{-g}}} \partial_{\mu}[\sqrt{-g} \partial^{\mu} \phi] = -e^{-2
\phi} (\partial_{\mu}a)^2  + {\frac {\alpha'} {4 g^2}} e^{\phi}{\cal{R}}^2
_{GB}\\ \end{equation}
\begin{equation}
{\frac {1} {\sqrt{-g}}} \partial_{\mu}[\sqrt{-g} e^{-2 \phi} \partial^{\mu} a]
= {\frac {\alpha'} {4g^2}}{\cal{R}} {\tilde{\cal{R}}}\\
\end{equation}
\begin{equation}
{\frac {1} {\sqrt{-g}}} \partial_{\mu}[\sqrt{-g} \partial^{\mu} \sigma] =
-e^{-2 \sigma}{({\partial} _{\mu} b)}^2 + {\frac{2 \alpha'} {3}}\left({\frac
{\delta \Delta} {\delta\sigma}} \right) {\cal{R}}^2 _{GB} + {\frac {2 \alpha'}
{3}} \left( {\frac {\delta \Theta} {\delta \sigma}}\right){\cal{R}}
{\tilde{\cal{R}}}\\
\end{equation}
\begin{equation}
{\frac {1} {\sqrt{-g}}} \partial_{\mu}[\sqrt{-g} e^{-2 \sigma}
\partial^{\mu}b] = {\frac{2 \alpha'} {3}}\left({\frac{\delta \Delta} {\delta
b}}
 \right) {\cal{R}}^2 _{GB} + {\frac {2 \alpha'}{3}} \left( {\frac {\delta
\Theta}  {\delta b}}\right){\cal{R}}{\tilde{\cal{R}}}\\
\end{equation}
and the equation\footnote{$\tilde{R}^{\mu \nu } _{\; \; \; \kappa \lambda}=
{\cal{\eta}}^{\mu \nu \rho \sigma} R_{\rho \sigma \kappa \lambda}$}
\begin{eqnarray}
& & R_{\mu \nu}-{\frac {1} {2}}g_{\mu \nu}R + \alpha'(g_{\mu \rho}
g_{\nu\lambda}
+ g_{\mu \lambda} g_{\nu \rho}) {\cal{\eta}}^{\kappa \lambda \alpha \beta}
D_{\gamma}({{\tilde{R}}^{\rho \gamma}} _{\;\;\;\alpha \beta} D_{\kappa}
f_1) - 8 \alpha' D_{\rho}(\tilde{R}_{\mu \; \; \nu} ^{\; \; \lambda
\; \; \rho}  D_{\lambda} f_2) =\nonumber\\
& &-{\frac{1} {2}}(\partial_{\mu} \phi)
(\partial_{\nu} \phi) + {\frac{1} {4}} g_{\mu \nu} (\partial_{\rho}\phi)^2 -
{\frac{e^{-2 \phi}} {2}} (\partial_{\mu}a) (\partial_{\nu}a) +{\frac
{1} {4}}g_{\mu \nu} e^{-2 \phi} (\partial_{\rho}a)^2\nonumber\\& & -{\frac{3}
{2}} (\partial_{\mu}\sigma) (\partial_{\nu} \sigma) + {\frac {3} {4}} g_{\mu
\nu}(\partial_{\rho}\sigma)^2 -{\frac {3} {2}}e^{-2
\sigma}(\partial_{\mu}b) (\partial_{\nu}b) + {\frac {3} {4}} e^{-2 \sigma}
g_{\mu \nu } (\partial_{\rho}b)^2
\end{eqnarray}
We  have introduced the functions
\begin{equation}
f_1 \equiv {\frac {e^{\phi}} {8 g^2}} + \Delta \quad , \quad f_2 \equiv {\frac
{a} {8 g^2}} + \Theta
\end{equation}

At this point we introduce the Kerr metric anticipating that it will continue
to be a solution to $O(\alpha')$
\begin{equation}
ds^2 = \left({\frac{{\rho}^2-2 M r} {{\rho}^2}}\right) dt^2 - {\frac{{\rho}^2}
{\Lambda}} dr^2 - {\rho}^2 d{\theta}^2 + {\frac {4 MrA{sin}^2\theta}
{{\rho}^2}}dt d\varphi-{\frac{{sin}^2\theta} {{\rho}^2}} {\Sigma}^2
d{\varphi}^2\\ \end{equation}
where  ${\rho}^2 \equiv r^2+A^2{cos}^2\theta$,$\Lambda \equiv r^2+A^2-2Mr$ and
${\Sigma}^2 \equiv (r^2+A^2)^2-\Lambda A^2{sin}^2\theta$.$A$ stands for the
angular momentum per unit mass.For this metric we can calculate
\begin{equation}
{\cal{R}}{\tilde{\cal{R}}} = {\frac
{192M^2Arcos\theta(3r^2-A^2{cos}^2\theta)(r^2-3A^2{cos}^2\theta)}
{(r^2+A^2{cos}^2\theta)^6}}\\
\end{equation}
%
\begin{equation}
{\cal{R}}^2 _{GB} = {\frac
{48M^2(r^2-A^2{cos}^2\theta)[(r^2+A^2{cos}^2\theta)^2-16r^2A^2{cos}^2\theta]}
{(r^2+A^2{cos}^2\theta)^6}}\\
\end{equation}

Since,as we declared in the introduction,we plan to determine solutions to
$O(\alpha')$ let us first obtain the zeroth order solutions for the scalar and
pseudoscalar fields.Introducing a rescaled axion field $\partial_{\mu}
\tilde{a} \equiv e^{-2 \phi} \partial_{\mu}a$ we can write the dilatonic-axion
equation of motion in the form
\begin{equation}
{\frac {\partial} {\partial r}}\left[(r^2-2Mr+A^2){\frac{\partial \tilde{a}}
{\partial r}}\right] + {\frac {1} {sin\theta}}{\frac{\partial} {\partial
\theta}} \left[sin\theta {\frac{\partial \tilde{a}} {\partial \theta}}\right] =
0\\
\end{equation}
It has a general solution of the form
\begin{equation}
\tilde{a} = \sum_{l=0}^{\infty} P_l (cos\theta)\left[A_l Q_l(z) + B_l
P_l(z)\right]\\
\end{equation}
where $z\equiv (r-M)/\sqrt{M^2-A^2}$. Imposing the black hole boundary
condition\footnote{$r_H=M+\sqrt{M^2-A^2}$} $ r \rightarrow r_H$  or  $
z\rightarrow 1$  forces us to require $ A_l =0 $, $\forall l$.On the other hand
requiring finiteness at  $r\rightarrow \infty$  or  $z \rightarrow \infty$
forces us to set $ B_l=0$ , $\forall l\geq1$. Thus, only the constant solution
$
\tilde{a}= B_o$ is possible.Using that,the dilaton equation reduces,to zeroth
order,to the form $D^2\phi=0$ which for the same reasons as in the case of the
axion $\tilde{a}$ leads us to the conclusion that to this order the dilaton is
a
constant. Following the same procedure for the modulus and its associated axion
we also arrive at constant zeroth order values.

In order to proceed and obtain the $O(\alpha')$ solutions we need the static
axisymmetric Green's function defined by the equation
\begin{equation}
{\frac {1} {\sqrt{-g}}} \partial_{\mu}\left[ \sqrt{-g} g^{\mu \nu}
\partial_{\nu}G(x-y)\right] = {\frac {\delta^{(3)}(x-y)} {\sqrt{-g}}}\\
\end{equation}
which for our metric $(13)$ becomes
\begin{equation}
{\frac {\partial} {\partial r}}\left[(r^2+A^2-2Mr){\frac {\partial G} {\partial
r}}\right]+ {\frac{1} {sin\theta}} {\frac {\partial}
{\partial \theta}}\left[sin\theta {\frac {\partial G} {\partial
\theta}}\right] = -\delta(r-r_o) \delta(cos\theta-cos{\theta}_o)
\delta(\varphi-{\varphi}_o)\\
\end{equation}
for a point source located at $r_o$ , $\theta_o$ , $ \varphi_o$.Demanding
finiteness at  $ r=r_H$ and at infinity we obtain
\begin{equation}
G(r,\theta,\varphi; r_o,\theta_o,\varphi_o) = \sum_{l=0}^{\infty} R_l (r,r_o)
P_l (cos\gamma)\\
\end{equation}
with
\begin{equation}
cos\gamma \equiv cos\theta cos\theta_o + sin\theta sin\theta_o
cos(\varphi-\varphi_o)\\ \end{equation}
and
\begin{eqnarray}
R_l  (r,r_o)& = & - {\frac {(2l+1)} { 4\pi \sqrt{M^2-A^2}}}  \left[ P_l
\left({\frac {(r_o-M)} {\sqrt{M^2-A^2}}} \right)  Q_l \left(
{\frac {(r-M)} {\sqrt{M^2-A^2}}} \right)  \theta(r-r_o) \right. \nonumber\\
& + & \left.  P_l\left( {\frac {(r-M)} {\sqrt{M^2-A^2}}} \right) Q_l
\left({\frac{(r_o-M)} {\sqrt{M^2-A^2}}}\right)  \theta( r_o-r) \right]
\end{eqnarray}

Using the Green's function we can write  the external dilaton solution as
\begin{equation}
\phi(r,\theta,\varphi) = \int_{r_H}^{\infty} dr_o \int_{-1}^1 dcos{\theta}_o
\int_{0}^{2\pi}d{\varphi}_o (r_o^2 + A^2 {cos}^2\theta_o)
G(r,\theta,\varphi;r_o,\theta_o,\varphi_o) {\cal{J}}(r_o,\theta_o,\varphi_o)
\end{equation}
where the source ${\cal{J}}$ is the right hand side of equation $(7)$.Similar
expressions hold for the rest of the scalar and pseudoscalar fields\footnote{$
\partial_{\mu}\tilde{b}=e^{-2\sigma} \partial_{\mu} b$}
$\sigma$,$\tilde{a}$,$\tilde{b}$ in terms of the corresponding  source
terms.It is straightforward but tedious to obtain the $O(\alpha')$ expressions
for these fields.At the same time ,since all scalar and pseudoscalar fields
have
non-constant parts of order $\alpha'$,the right hand side of equation $(11)$ is
$O({\alpha'}^2)$ and thus the gravitational part of the solution is the
standard
Kerr metric $(13)$.The $O(\alpha')$ fields are\newpage
\begin{eqnarray}
\phi(r,\theta)& =& \phi_o - {\frac { \alpha' e^{\phi_o} } {g^2}}  \left[
{\frac
{1} { A^2}} ln\left({\frac {r-M+\sqrt{M^2-A^2}} {\sqrt{A^2+r^2}}}\right) +
{\frac {M r} {(A^2+r^2)^2}} \right. \nonumber\\&+&\left.
 {\frac {(2 A^2-M^2)} {2 A^3 M}}\left(
{\frac{\pi} {2}} - Arctan({\frac {r} {A}})\right) + {\frac {2 A^2+M r} {2 A^2
 (A^2+r^2)}}  \right]P_o(cos\theta) +...
\end{eqnarray}
\\

\begin{eqnarray}
\tilde{a}(r,\theta) & = & \tilde{a}_o- {\frac {6\alpha'A} {g^2 (M^2-A^2)}}
\left\{
  (r-M) \left[ {\frac {(A^2- M^2)} {  A^4}}
ln \left( {\frac {r-M+\sqrt{M^2-A^2}} {\sqrt{A^2+r^2}}} \right)\right.
\right. \nonumber \\& +&\left.  \left.{\frac {2 A^2+Mr-M^2} {2 A^2
(A^2+r^2)}} +{\frac {(A^2-M^2)} { A^3 M}} \left( {\frac {\pi} {2}} -
Arctan({\frac {r} {A}})\right) \right] + {\frac {A^2 \sqrt{M^2-A^2}} { M
r_H ^3}}\right. \nonumber \\& +&  \left. {\frac {(3 M-4 r_H)} {2
 r_H ^2}} +{\frac {M (M^2+r^2)} { (A^2+r^2)^2}} \right\}P_1(cos\theta) +...
\end{eqnarray}
\\

\begin{eqnarray}
\sigma(r,\theta) & = &\sigma_o- {\frac {8\alpha'} {3}} \left( {\frac{\partial
\Delta} {\partial \sigma}} \right)_{\sigma_o,b_o} \left[{\frac {1}
{ A^2}} ln\left({\frac {r-M+\sqrt{M^2-A^2}} {\sqrt{A^2+r^2}}}\right) +
{\frac {M r} {(A^2+r^2)^2}} \right. \nonumber\\&+&\left.
 {\frac {(2 A^2-M^2)} {2 A^3 M}}\left(
{\frac{\pi} {2}} - Arctan({\frac {r} {A}})\right) + {\frac {2 A^2+M r} {2 A^2
 (A^2+r^2)}}  \right]P_o(cos\theta) \nonumber \\
 & - &{\frac {2\alpha'} {3}}
\left( {\frac{\partial \Theta}  {\partial \sigma}} \right)_{\sigma_o,b_o}{\frac
{24 A} { (M^2-A^2)}} \left\{
  (r-M) \left[ {\frac {(A^2- M^2)} {  A^4}}
ln \left( {\frac {r-M+\sqrt{M^2-A^2}} {\sqrt{A^2+r^2}}} \right)\right.
\right. \nonumber \\& +&\left.  \left.{\frac {2 A^2+Mr-M^2} {2 A^2
(A^2+r^2)}} +{\frac {(A^2-M^2)} { A^3 M}} \left( {\frac {\pi} {2}} -
Arctan({\frac {r} {A}})\right) \right] + {\frac {A^2 \sqrt{M^2-A^2}} { M
r_H ^3}}\right. \nonumber \\& +&  \left. {\frac {(3 M-4 r_H)} {2
 r_H ^2}} +{\frac {M (M^2+r^2)} { (A^2+r^2)^2}} \right\}P_1(cos\theta) +...
\end{eqnarray}
\\

\begin{eqnarray}
\tilde{b}(r,\theta) & = &\tilde{b}_o- {\frac {8\alpha'} {3}} \left(
{\frac{\partial \Delta} {\partial b}} \right)_{\sigma_o,b_o} \left[{\frac {1}
{ A^2}} ln\left({\frac {r-M+\sqrt{M^2-A^2}} {\sqrt{A^2+r^2}}}\right) +
{\frac {M r} {(A^2+r^2)^2}} \right. \nonumber\\&+&\left.
 {\frac {(2 A^2-M^2)} {2 A^3 M}}\left(
{\frac{\pi} {2}} - Arctan({\frac {r} {A}})\right) + {\frac {2 A^2+M r} {2 A^2
 (A^2+r^2)}}  \right]P_o(cos\theta) \nonumber \\
&-&{\frac {2\alpha'} {3}}
\left( {\frac{\partial \Theta}  {\partial b}} \right)_{\sigma_o,b_o}{\frac {24
A}
{ (M^2-A^2)}} \left\{
  (r-M) \left[ {\frac {(A^2- M^2)} {  A^4}}
ln \left( {\frac {r-M+\sqrt{M^2-A^2}} {\sqrt{A^2+r^2}}} \right)\right.
\right. \nonumber \\& +&\left.  \left.{\frac {2 A^2+Mr-M^2} {2 A^2
(A^2+r^2)}} +{\frac {(A^2-M^2)} { A^3 M}} \left( {\frac {\pi} {2}} -
Arctan({\frac {r} {A}})\right) \right] + {\frac {A^2 \sqrt{M^2-A^2}} { M
r_H ^3}}\right. \nonumber \\& +&  \left. {\frac {(3 M-4 r_H)} {2
 r_H ^2}} +{\frac {M (M^2+r^2)} { (A^2+r^2)^2}} \right\}P_1(cos\theta) +...
\end{eqnarray}

  The leading modulus and $\tilde{b}$-hair behaviour is that of a monopole term
analogous to the dilaton.This is evident from the slow rotation limit of the
dilaton solution
\begin{eqnarray}
\phi(r,\theta)& = &\phi_o + {\frac {\alpha' e^{\phi_o}} {4 g^2}} \left[-{\frac
{2} {M r}} (1+{\frac {M} {r}} + {\frac {4 M^2} {3 r^2}}) + {\frac {A^2} {2 M^3
r}} ( {\frac {1} {2}}+ {\frac {M} {r}} + {\frac {12 M^2} {3 r^2}}\right.
\nonumber\\ & +& \left.{\frac {6 M^3} {r^3}} + {\frac {64 M^4} {5 r^4}} )
+...\right] P_o (cos\theta)+... \end{eqnarray}
Note that the coefficient functions$^{\cite{Antoniadis}}$ $\Delta$ and
$\Theta$ are such that they have an extremum at the self-dual point
$\sigma_o=\tilde{b}_o=0$.Perturbing around the self-dual solution leads to
vanishing modulus hair to $O(\alpha')$.The infinite continuum of non-zero
$\sigma_o$,$\tilde{b}_o$ values allows for the non-vanishing modulus and
$\tilde{b}$-axion hair given by $(26)$ and $(27)$.

Although the existence of non-trivial dilaton,moduli and axion fields outside
a Kerr black hole seems to violate the letter of the ``no-hair theorem" it
does not violate the spirit since the solution is uniquelly characterized by
mass and angular momentum.In the terminology introduced by
S.Coleman,J.Preskill and F.Wilczek $^{\cite{Coleman}}$ the external moduli
and dilaton hair are examples of ``secondary " hair.
\\

\vspace*{1in}
\noindent
{\bf Acknowledgments}\\ \\
We thank I.Antoniadis,N.Mavromatos and J.Rizos for illuminating discussions.



\begin {thebibliography}{99}
\bibitem{Green}For a review,see :M.Green , J.Schwarz and  E.Witten,
``Superstring Theory " (Cambridge U.P., Cambridge,1987).
\bibitem{Callan} C.G.Callan, D.Friedan, E.J.Martinec and M.J.Perry, Nucl.Phys.
B262 (1985)543 ; B278 (1986)78.\\
E.S.Fradkin and A.A.Tseytlin, Phys.Lett.B158(1985)316; Nucl.Phys.B262(1985)1\\
A.Sen,Phys.Rev. D32 (1985)2142; Phys.Rev.Lett.55(1985)1846.
\bibitem{Wald} R.M.Wald ,General Relativity (University of Chicago Press,
Chicago,1984) and references therein.
\bibitem{Israel} W.Israel,Phys.Rev.164(1967)331;for a review,\\
in ``General Relativity, an Einstein Centennary Survey"
edit.S.W.Hawking and W.Israel(Cambridge U.P.,Cambridge,1979).\\
B.Carter,Phys.Rev.Lett.26(1971)331\\
R.Price,Phys.Rev.D5(1972)2419,2439\\
J.D.Bekenstein,Phys.Rev.D5(1972)1239,2403\\
C.Teitelboim,Phys.Rev.D5(1972)2941\\
D.C.Robinson,Phys.Rev.Lett.34(1975)405\\
R.M.Wald,Phys.Rev.Lett.26(1971)1653\\
E.D.Fackerell and J.R.Ipser,Phys.Rev.D5(1972)2455\\
S.L.Adler and R.B.Pearson,Phys.Rev.D18(1978)2798
\bibitem{Gibbons} G.W.Gibbons,Nucl.Phys.207(1982)337\\
G.W.Gibbons and K.Maeda,Nucl.Phys.B298(1988)741
\bibitem{Perry} C.G.Callan,R.C.Myers and M.J.Perry,Nucl.Phys.B311(1988/89)673
\bibitem{Campbell} B.Campbell,M.Duncan,N.Kaloper and
K.A.Olive,Phys.Lett.B251(1990)34
\bibitem{Garfinkle} D.Garfinkle,G.T.Horowitz and
A.Strominger,Phys.Rev.D43(1991)3140
\bibitem{Kaloper} B.Campbell,N.Kaloper and K.A.Olive,Phys.Lett.B263(1991)364
\bibitem{Shapere} A.Shapere,S.Trivedi and F.Wilczek,Mod.Phys.Lett.A6(1991)2677
\bibitem{Olive} B.Campbell,N.Kaloper and K.A.Olive,Phys.Lett.B285(1992)199
\bibitem{Antoniadis} I.Antoniadis,E.Gava and
K.S.Narain,Phys.Lett.B283(1992)209;\\ Nucl.Phys.B393(1992)93
\bibitem{Tamvakis} I.Antoniadis,J.Rizos and K.Tamvakis,Nucl.Phys.B415(1994)497
\bibitem{Coleman} S.Coleman,J.Preskill and F.Wilczek,Nucl.Phys.B378(1992)175
\end{thebibliography}
\end{document}